\documentclass{JINST}

\RequirePackage{xspace}

\def\mvfc     {\ensuremath{{\rm \,mV\!\!/fC}}\xspace}

\def\adufc    {\ensuremath{{\rm \,ADU\!/fC}}\xspace}

\def\volt     {\ensuremath{{\rm \,V}}\xspace}
\def\fc   	  {\ensuremath{{\rm \,fC}}\xspace}

\def\voltcm		{\ensuremath{{\rm \,V\!\cdot cm^{-1}}\xspace}}

\def\lh				{\ensuremath{{\rm \,l/h}}\xspace}

\def\mbar     {\ensuremath{{\rm \,mbar}}\xspace}
\def\mbarinv     {\ensuremath{{\rm \,mbar^{-1}}}\xspace}

\def\degk 		{\ensuremath{\rm \,K}\xspace}
\def\degkinv 		{\ensuremath{\rm \,K^{-1}}\xspace}

\def\pad     	{\ensuremath{{\rm \,pad}}\xspace}
\def\adu	    {\ensuremath{{\rm \,ADU}}\xspace}

\def\cm 		  {\ensuremath{{\rm \,cm}}\xspace}
\def\mm 		  {\ensuremath{{\rm \,mm}}\xspace}
\def\mum			{\ensuremath{{\,\mu\rm m}}\xspace}

\def\ma			  {\ensuremath{{\rm \,m}^2}\xspace}
\def\cma  		{\ensuremath{{\rm \,cm}^2}\xspace}
\def\mma  		{\ensuremath{{\rm \,mm}^2}\xspace}

\def\muminv			{\ensuremath{{\,\mu\rm m^{-1}}}\xspace}

\def\hour				{\ensuremath{\rm {\,h}}\xspace}
\def\sec				{\ensuremath{\rm {\,s}}\xspace}      
\def\mus        {\ensuremath{\,\mu{\rm s}}\xspace}   

\def\tev				{\ensuremath{\mathrm{\,Te\kern -0.1em V}}\xspace}
\def\gev				{\ensuremath{\mathrm{\,Ge\kern -0.1em V}}\xspace}
\def\mev				{\ensuremath{\mathrm{\,Me\kern -0.1em V}}\xspace}
\def\kev				{\ensuremath{\mathrm{\,ke\kern -0.1em V}}\xspace}
\def\ev					{\ensuremath{\mathrm{\,e\kern -0.1em V}}\xspace}

\def\gevc				{\ensuremath{{\mathrm{\,Ge\kern -0.1em V\!/}c}}\xspace}
\def\mevc				{\ensuremath{{\mathrm{\,Me\kern -0.1em V\!/}c}}\xspace}

\def\gevcc			{\ensuremath{{\mathrm{\,Ge\kern -0.1em V\!/}c^2}}\xspace}
\def\mevcc			{\ensuremath{{\mathrm{\,Me\kern -0.1em V\!/}c^2}}\xspace}

\def\en				  {\ensuremath{e^-}\xspace} 

\def\epem       {\ensuremath{e^+e^-}\xspace}

\def\w      {\ensuremath{W}\xspace}

\def\z      {\ensuremath{Z^0}\xspace}

\def\ariso {Ar/iC$_4$H$_{10}$\xspace}
\def\arisoProp {Ar/iC$_4$H$_{10}$ (95/5)\xspace}
\def\arco {Ar/CO$_2$\xspace}
\def\arcoProp {Ar/CO$_2$ (80/20)\xspace}

\title{MICROMEGAS chambers for hadronic calorimetry at a future linear collider}

\author{C. Adloff$^a$, D. Atti\'e$^b$, J. Blaha$^a$, S. Cap$^a$, M. Chefdeville$^a$, P. Colas$^b$, A. Dalmaz$^a$, C. Drancourt$^a$, A. Espargili\`{e}re$^a$\thanks{Corresponding author.}~, R. Gaglione$^a$, R. Gallet$^a$, N. Geffroy$^a$, I. Giomataris$^b$, J. Jaquemier$^a$, Y. Karyotakis$^a$, F. Peltier$^a$, J. Prast$^a$, G. Vouters$^a$

\\
\llap{$^a$}Laboratoire d'Annecy-le-Vieux de Physique des Particules (LAPP), Université de Savoie, CNRS/IN2P3,\\
  9 Chemin de Bellevue, F-74941 Annecy-le-Vieux, France\\
\llap{$^b$} Institut de recherche sur les lois fondamentales de l'Univers (IRFU), CEA, Saclay, France\\
  
  E-mail: \email{ambroise.espargiliere@lapp.in2p3.fr}
	}

\abstract{
Prototypes of MICROMEGAS chambers, using bulk technology and analog readout, with 1$\times$1$\cma$ readout segmentation have been built and tested. 

Measurements in Ar/iC$_4$H$_{10}$ (95/5) and Ar/CO$_2$ (80/20) are reported. The dependency of the prototypes gas gain versus pressure, gas temperature and amplification gap thickness variations has been measured with an $^{55}$Fe source and a method for temperature and pressure correction of data is presented.

A stack of four chambers has been tested in 200$\gevc$ and 7$\gevc$ muon and pion beams respectively. 
Measurements of response uniformity, detection efficiency and hit multiplicity are reported.

A bulk MICROMEGAS prototype with embedded digital readout electronics has been assembled and tested.
The chamber's layout and first results are presented.
}


\begin{document}

\section{Introduction}
\begin{sloppypar}
Future linear $\epem$ colliders at Terascale energies, like the International Linear Collider (ILC) or the Compact LInear Collider (CLIC), will be the probes for new physics. 
Depending on Large Hadron Collider (LHC) results they will be able to get unprecedented measurements on Higgs physics but also on Super Symmetry and Standard Model extensions \cite{ILCphys}. 

To obtain the unequaled jet energy resolution required to separate $\w$ and $\z$ induced jets ($\sim$30\%$/\sqrt{E}$), one of the most promising analysis technique is based on Particle Flow Algorithms (PFA) \cite{PFAbrient, PFA}. 
Jet energy is measured by combining the measurements of the track momentum from charged particles and of the calorimetric energy from neutrals.
With this technique, the jet energy resolution is dominated by the error in hit assignment to clusters (so-called PFA's confusion term), it is therefore mandatory to discriminate between charged and neutral clusters. This calls for highly segmented calorimeters with the the capability to produce narrow showers. 

The high segmentation together with the large area to be instrumented (\emph{e.g.} $\sim$3000$\ma$ for the SiD hadronic calorimeter \cite{SIDLOI}) leads to a dramatic increase of the readout channel number and thus of the amount of data to handle and store. 
This can be balanced by using a Digital Hadronic CALorimeter (DHCAL) counting the number of hits rather than measuring the deposited energy. 
In a sampling DHCAL, a few options concerning the active layer are considered: scintillators tiles, GEMs, RPCs and MICROMEGAS \cite{ILCCAL, SIDLOI, ILDLOI}. 

The MICRO-MEsh GAseous Structure (MICROMEGAS) was introduced in 1996 \cite{bibMicro1} as a fast signal, position-sensitive, radiation hard gaseous detector. 
MICROMEGAS consists of a conductive mesh held a few tens of micrometers above a segmented anode plane,  defining the amplification gap, surmounted by a cathode defining the drift gap. An incident charged particle crossing the drift gap ionizes the gas. Using suitable voltage settings, the ionization electrons drift to the mesh, enter the high amplification field region where they produce further ionizations in cascade. The motion of charges in this region induces a signal on the mesh and on the anode plane. 

Large area MICROMEGAS based detectors have already been produced (up to 40$\times$40$\cma$ \cite{T2K}), using the so-called bulk MICROMEGAS technique. Such a bulk MICROMEGAS is made by lamination of photoresistive film layers on the readout Printed Circuit Board (PCB), strongly encapsulating a stretched micro-woven mesh at a fixed distance from the PCB, and forming spacer pillars after photolithography \cite{bulk}. 
With this manufacturing technique, a bulk MICROMEGAS with anode pad on one side of the PCB and embedded readout electronics on the back side offers a compact and robust detector that can be produced by industry. MICROMEGAS is therefore a very appealing possibility to equip a DHCAL. 

The present work was carried out in the framework of the Calorimeters for ILC Experiments collaboration (CALICE) \cite{CALICE} with the aim to test the bulk MICROMEGAS technology for DHCAL. 
 For extensive characterization, measurements were performed with bulk MICROMEGAS prototypes with external analog readout electronics boards equipped with GASSIPLEX chips \cite{GASSIPLEX}. 
 The feasibility of a compact detector has been studied by building a bulk MICROMEGAS with an embedded digital readout chip called DIRAC \cite{DIRAC}. 
 
 
\end{sloppypar}


\section{Experimental Setup}

\subsection{Analog Readout Prototypes}
\label{proto}
\begin{figure}[h]
	\centering
		\includegraphics[width=0.53\textwidth]{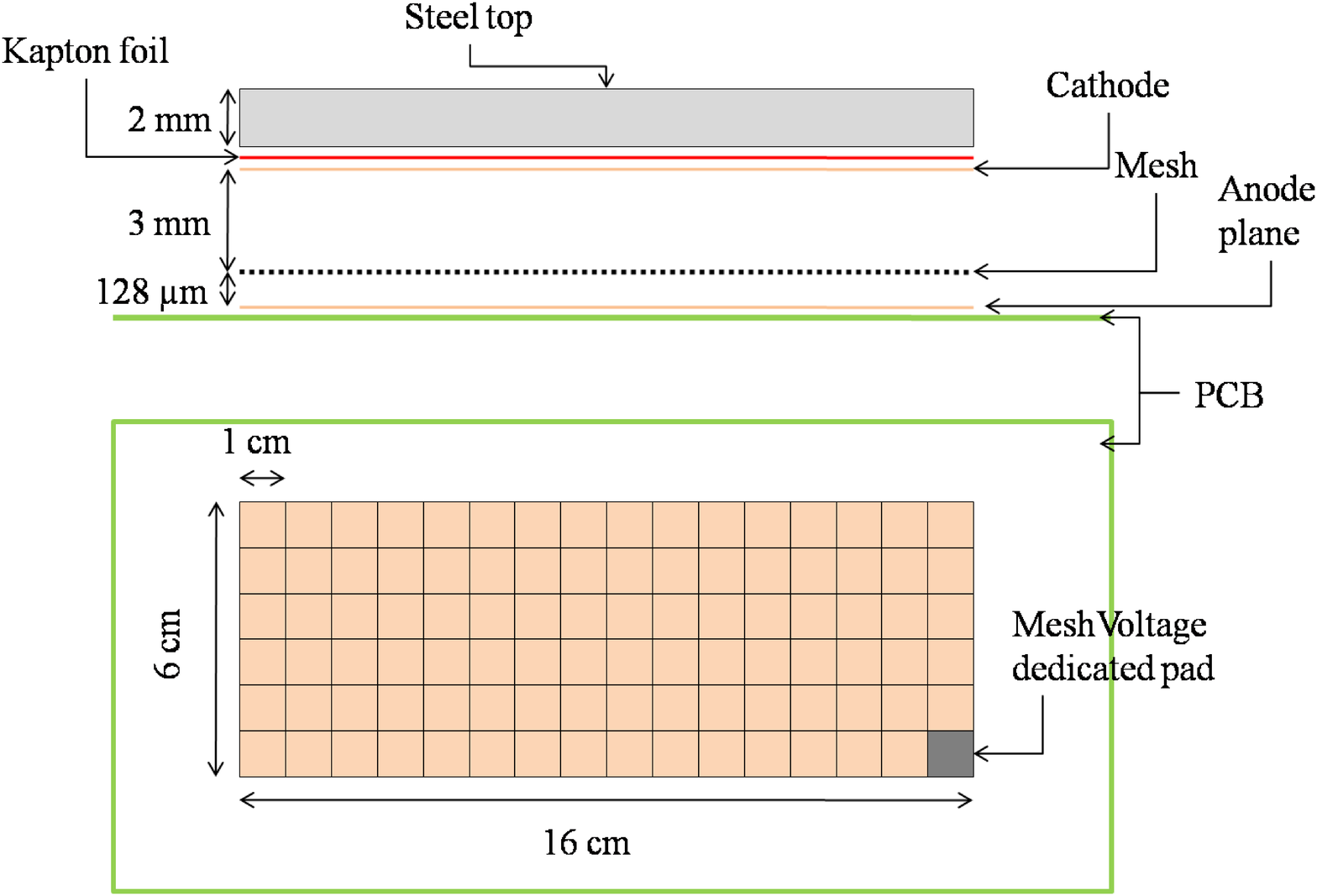}
		\includegraphics[width=0.46\textwidth]{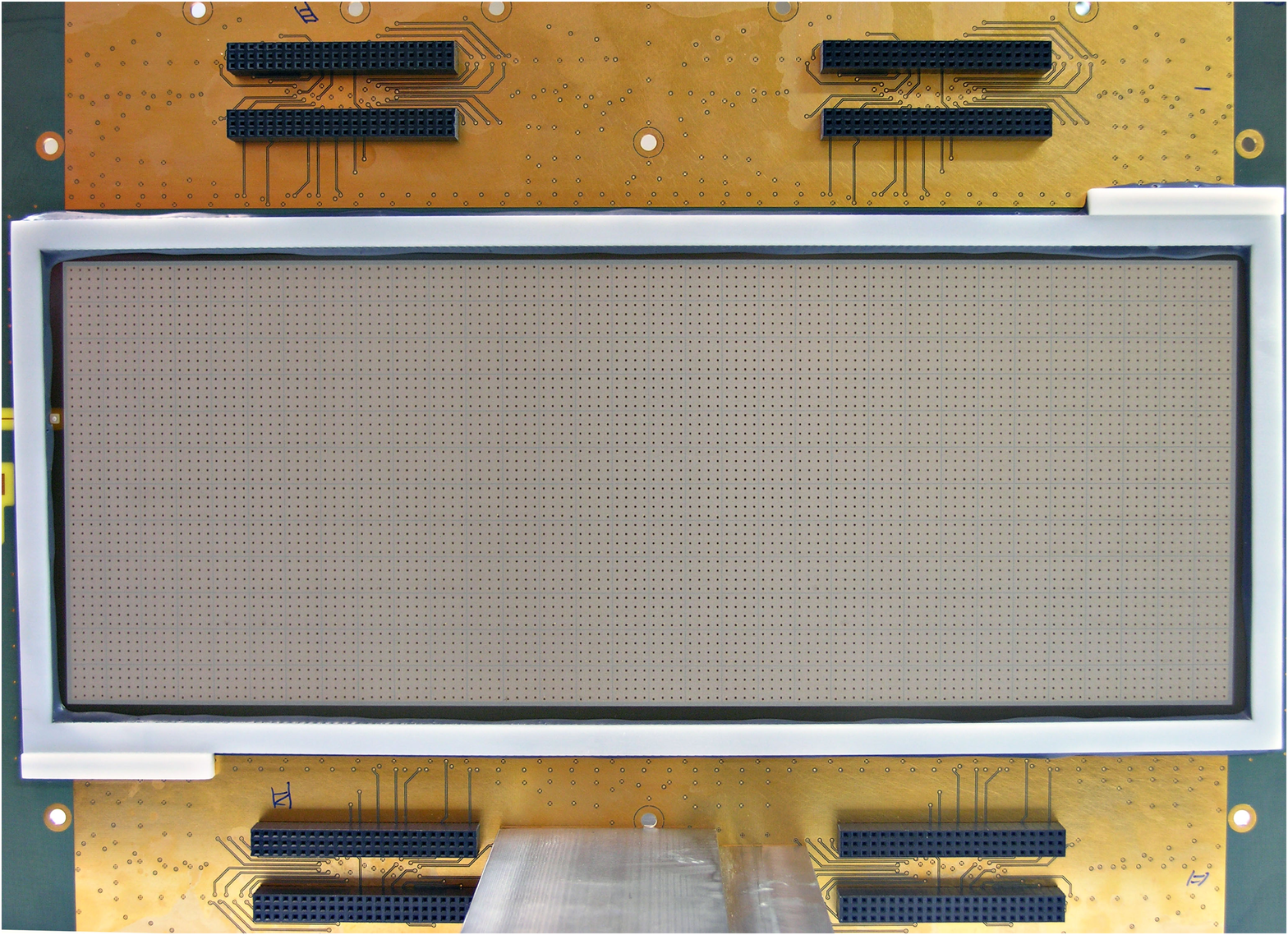}
	\caption{
		Analog readout prototypes. The left drawings give CH0 to CH2 dimensions but can be straightforwardly extended to CH3. The right picture shows the bulk, the resin frame and the electronic board connectors of CH3 before its lid was glued. 
	}
	\label{protoscheme}
\end{figure}

Each prototype consists of a bulk MICROMEGAS chamber with a 3$\mm$ drift gap and a 128$\mum$ amplification gap. 
The drift cathode is a 5$\mum$ thick copper foil fixed on a 75$\mum$ thick Kapton film. The whole is glued on a 2$\mm$ thick steel plate, forming the device's lid. 
The steel cover plate is part of the absorber and therefore would not contribute to the HCAL active layer's thickness. The 3$\mm$ drift gap is ensured by a 3$\mm$ thick resin frame enclosing the chamber and providing the gas inlet and outlet (see photography on figure \ref{protoscheme}, right). 

The bulk uses an industrial micro woven stainless steel mesh made of 30$\mum$ diameter wires interwoven at a pitch of 80$\mum$.  The mesh is held by 128$\mum$ high, 300$\mum$ diameter pillars laid out on a square lattice with a 2$\mm$ pitch. The anode plane consists of 0.98$\times$0.98$\cma$ pads spaced every 200$\mum$ lying on the detector's PCB. 
The PCB is a 4 layers class 4, 1.6$\mm$ thick. 
The 1\cma pattern made of a pad and the free space around will be denoted hereafter as a 1$\pad$ area where '\pad' will be the unit area symbol. 

Four prototypes with analog readout have been built. Three of them have a 6$\times$16$\cma$ active area (96\pad) and the last one is four times larger with a 12$\times$32$\cma$ active area (384\pad). The mesh voltage of the small chambers is supplied through a dedicated pad while a 4$\mma$ sidelong mesh voltage pad is used for the large chamber (see figure \ref{protoscheme}). 
In the following, the three small chambers will be denoted CH0, CH1 and CH2 and the large one, CH3. 

\subsection{Readout System}

\label{readout}
The anode analog readout was provided by 16-channel ASICs called GASSIPLEX \cite{GASSIPLEX}. 
Boards\footnote{CEA DAPNIA Board N$^\circ$613V, 96 channels, 6 GASSIPLEX chips 0.7 v3.}, each equipped with 6 GASSIPLEX chips, were mounted on the side of the chambers. 
One board was used for each of the small chambers and four for the large one.

GASSIPLEX chips, when triggered, gather the signal from every channel and build one single multiplexed differential output with a nominal conversion factor of 3.6\mvfc and a peaking time of 1.2\mus. 
The multiplexed signal from GASSIPLEX boards was digitized by CAEN V550 10-bit ADCs (VME modules) sequenced by a CAEN V551B C-RAMS sequencer VME module. Data were then collected by the computer through an optical VME/PCI bridge. 
A Labview based software, called CENTAURE \cite{CENTAURE}, was used for online monitoring and data recording. 
A very similar readout system was used for the CAST experiment and is described in \cite{CAST}. 
The global conversion ratio of the GASSIPLEX-based readout chain was measured to be 4.69$\pm$0.25$\adufc$ (Analog to Digital Unit). Its r.m.s. over all 672 channels is 2.5\%.

For the X-ray study, only one chamber was used with a different readout system, based on the mesh signal.  
The mesh signal was read out by an ORTEC 142C charge preamplifier linked to its corresponding amplifier/shaper.
The calibration constant of the mesh readout chain was precisely measured to be 2.199$\pm$0.026$\adufc$.

A detailed note about the calibration of both readout chains is available in \cite{Ren's}.  


%
\label{calib}
%


\section{X-ray Study}

X-ray tests using a 5.9\kev $^{55}$Fe source have been performed to measure the basic performance of the prototypes.  
For each prototype, the steel lid is drilled on a few locations to allow X-ray injection through the cathode and Kapton foil.  
The electron collection efficiency and the gas gain in Ar/iC$_4$H$_{10}$ (95/5) and Ar/CO$_2$ (80/20) were deduced from the $^{55}$Fe K$_{\alpha}$ photopeak value given by a fit of of three gaussian functions to the $^{55}$Fe spectrum (figure \ref{CollectionEff}, left). 
Gain measurements were used to predict gain dependency versus pressure, temperature and amplification gap thickness variations. Those predictions are confronted to direct measurements in section \ref{envstudy}.

\subsection{Electron Collection Efficiency}
\label{EFFCOLL}
\begin{figure}[h]
	\centering
	\includegraphics[width=0.46\textwidth]{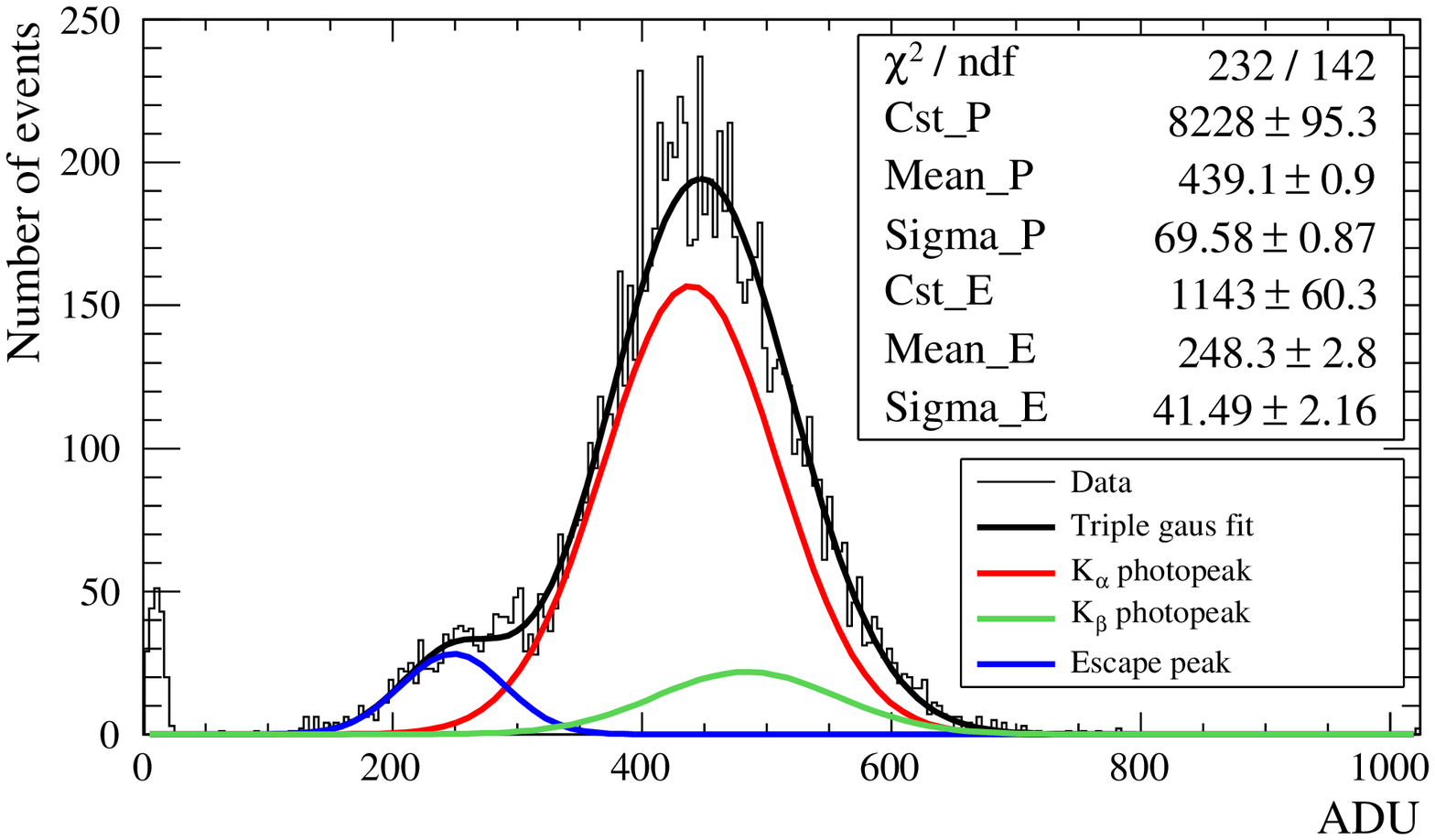}
		\includegraphics[width=0.53\textwidth]{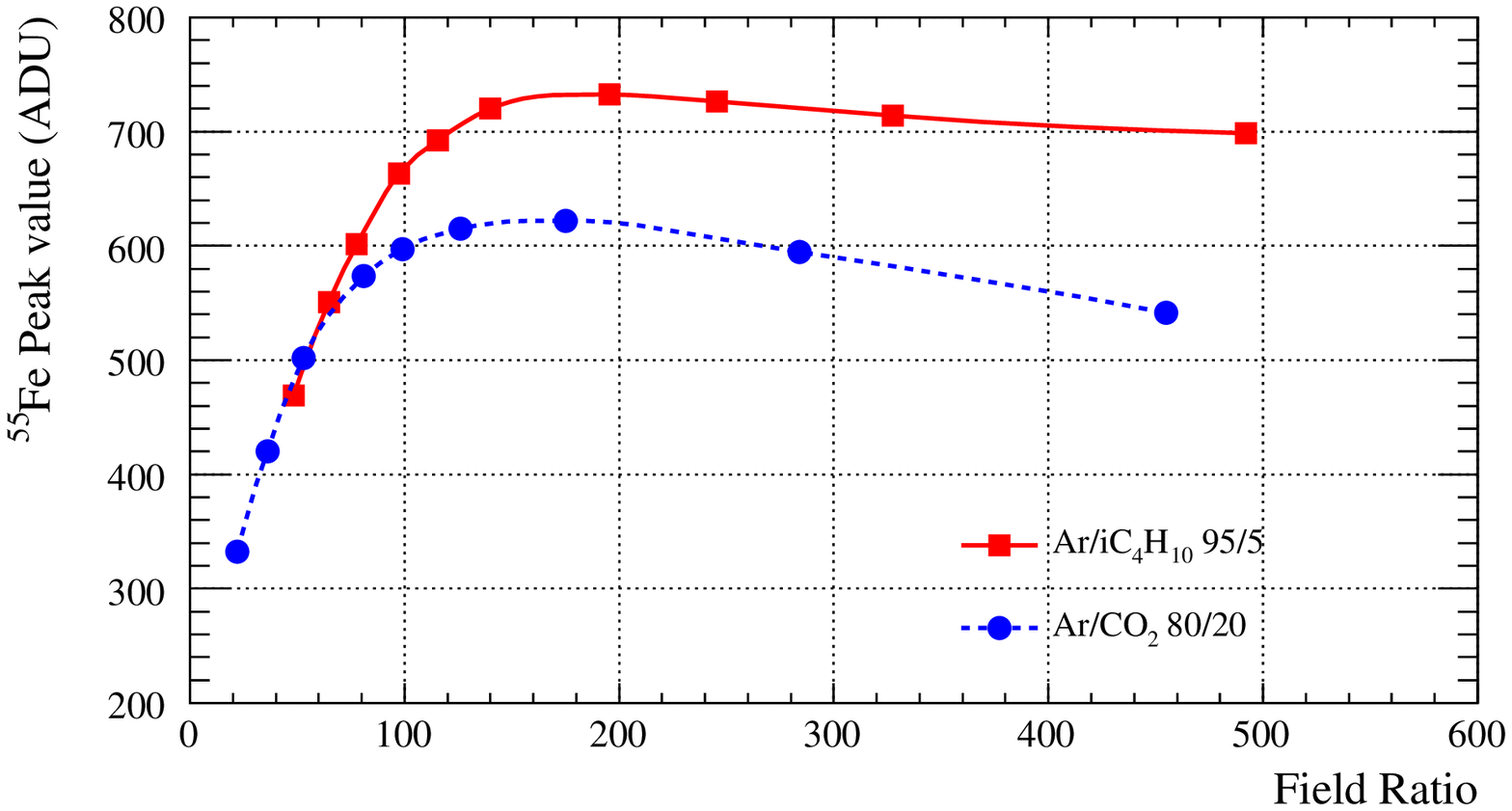}
	\caption{$^{55}$Fe spectrum with gaussian fits of the two photopeaks and of the escape peak (left). $^{55}$Fe peak value variation versus field ratio (right).}
	\label{CollectionEff}
\end{figure}

The ratio between amplification and drift electric fields affects the mesh transparency to electrons (or collection efficiency) by contracting the field lines so that the electrons are mostly driven through the center of the mesh's holes and reach the amplification gap. Figure \ref{CollectionEff} (right) shows the variation of the $^{55}$Fe peak value versus the field ratio in Ar/iC$_4$H$_{10}$ (95/5) and Ar/CO$_2$ (80/20). The amplification field was kept constant while the drift field varied to set the field ratio. The mesh voltage was 420\volt in Ar/iC$_4$H$_{10}$ (95/5) and 570\volt in Ar/CO$_2$ (80/20). 

The curves displayed on figure \ref{CollectionEff} (right) show a maximum at a field ratio of about 150 -- 200 for both gas mixtures. The measurements reported in the following were performed at ratios within this range.

 A possible explanation for the decline at high field ratio observed in figure \ref{CollectionEff} (right) is the attachment of some primary electrons in the drift region by electronegative impurities (\emph{e.g.} oxygen, water vapor). For a constant amplification field, a higher ratio means a lower drift field and consequently the primary electrons tend to have less energy. Since the attachment cross section of some impurities peaks at low energy (\emph{e.g.} $\approx$ 0.1\ev for oxygen \cite{magboltz}), a lower drift field can lead to a higher attachment probability.


\subsection{Gas Gain}
\begin{figure}[h]
	\centering
		\includegraphics[width=0.49\textwidth]{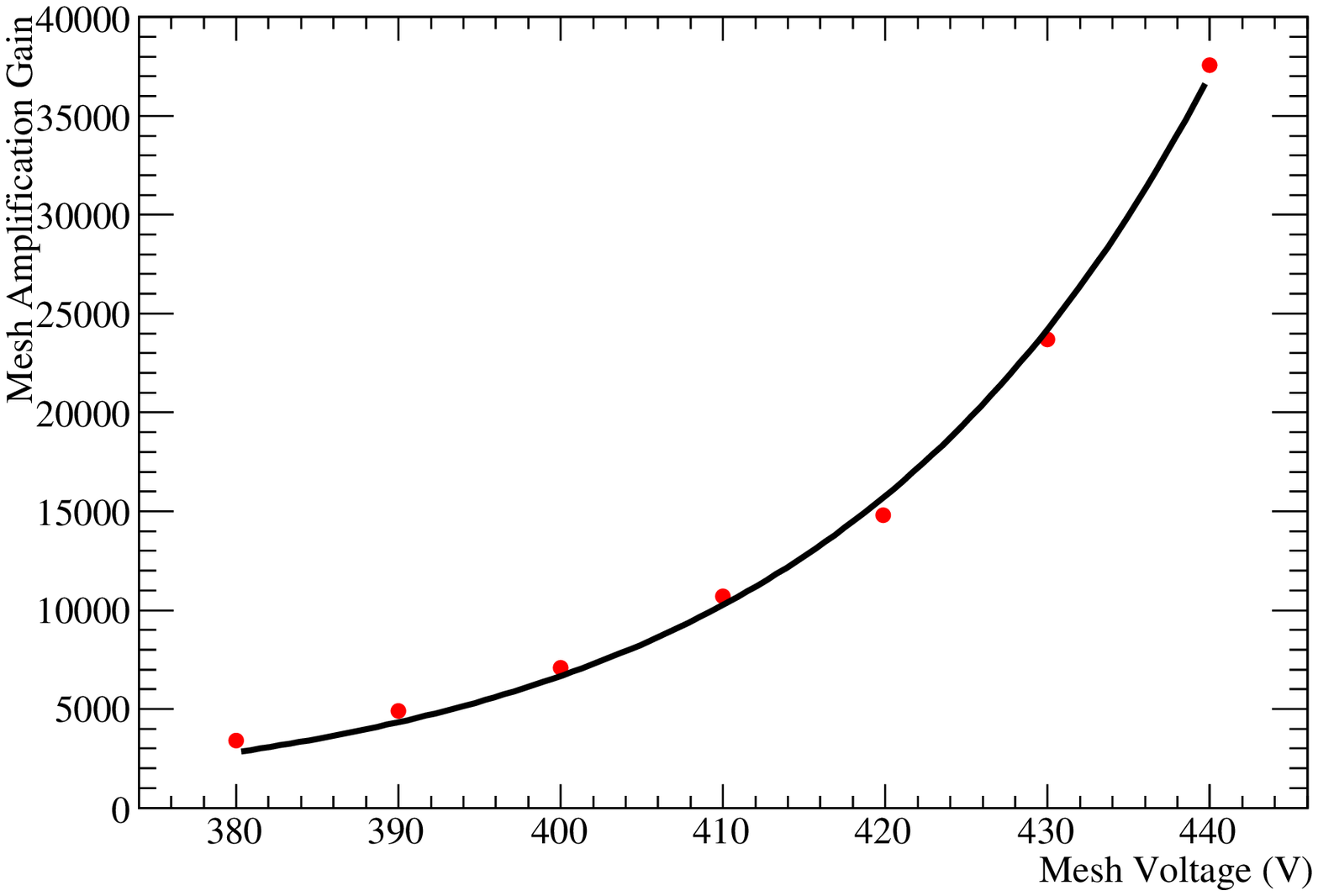}
		\includegraphics[width=0.49\textwidth]{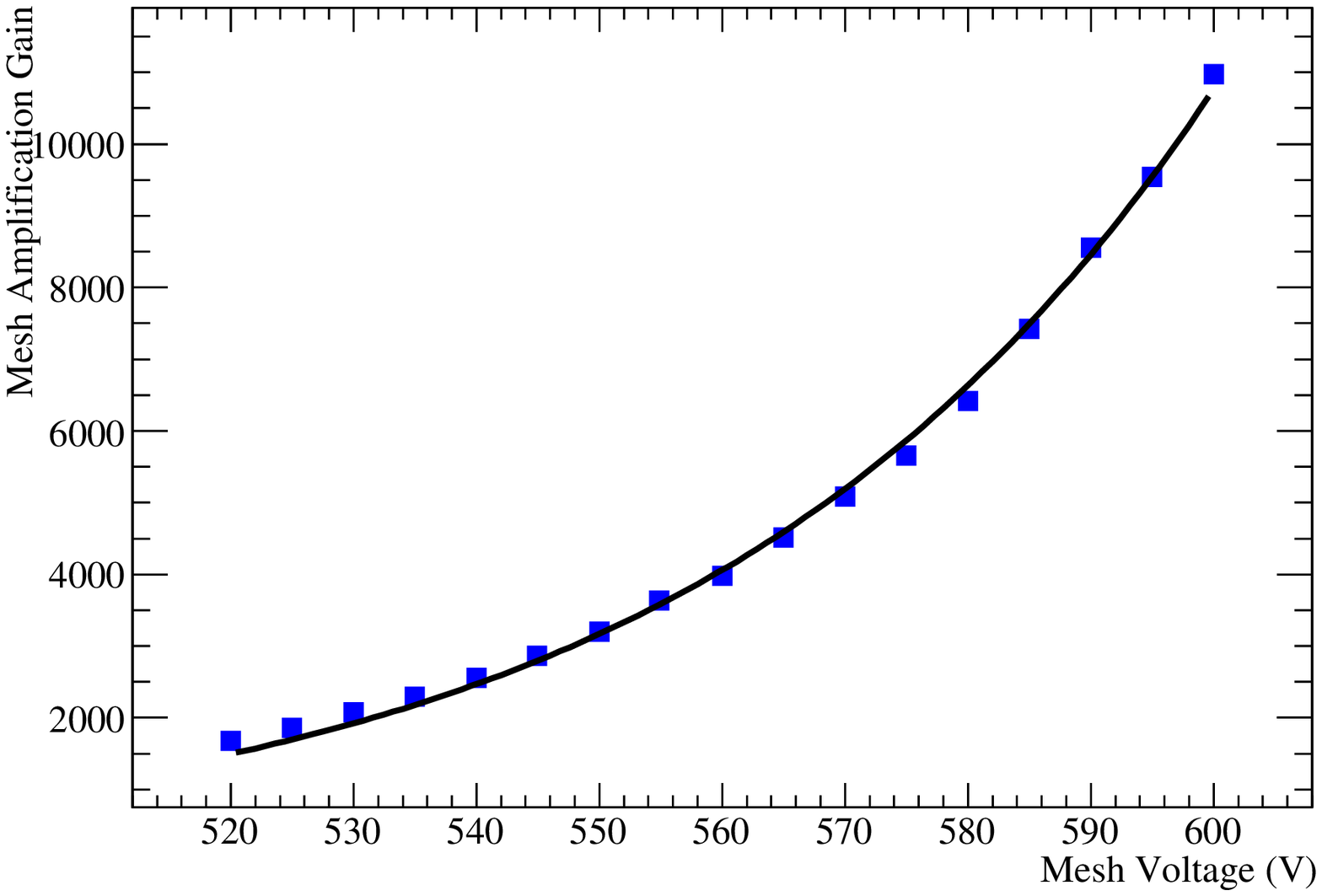}
	\caption{Gas gain versus mesh voltage fitted with the gas gain formula in \arisoProp (left) and in \arcoProp (right).}
	\label{GAINCURVES}
\end{figure}

The amplification gap gain, so-called gas gain, is determined through a fit of three gaussian functions to the $^{55}$Fe spectrum (figure \ref{CollectionEff}, left) assuming 230 and 209 primary electrons in the Ar/iC$_4$H$_{10}$ (95/5) and Ar/CO$_2$ (80/20) mixtures respectively and using the mesh readout calibration constant (2.19\adufc).
Keeping the drift field at 150\voltcm, a set of measurements at 954$\mbar$ in the \ariso mixture, 961\mbar in \arco and at a temperature of 293$\degk$ with various voltage settings gave the gain curves displayed in figure \ref{GAINCURVES}.

A much higher maximum gain (4$\cdot$10$^4$) can be achieved in \ariso than in \arco. Also the mesh voltage, for a given gain, is 180\volt lower in \ariso.

\subsection{Method for Pressure and Temperature Correction}
\subsubsection{Gas Gain Model}
\label{basics}
The gas gain of the chambers is given by the exponential of the average number of primary ionizations from a single avalanche initiating electron. In a uniform field, this number is given by the first Townsend coefficient, denoted $\alpha$, multiplied by the amplification gap thickness, denoted $g$: 
\begin{equation}
	\label{G0}
	G = e^{\alpha\cdot g} {\rm \,.}
\end{equation}
The Townsend coefficient can be parameterized by the Rose and Korff formula \cite{sauli}:
\begin{equation}
	\label{RK}
	\alpha = n\cdot A_0 e^{-B_0n/E} {\rm \,,}
\end{equation}
where $A_0$ and $B_0$ are constants that depends on the gas mixture, $E$ is the electric field and $n$ the gas number density. 
Using the ideal gas law to express $n$ and combining equations \ref{G0} and \ref{RK}, one obtain:
\begin{equation}
\label{G}
G=exp\left(\frac{APg}{T}\cdot exp\left(-\frac{BPg}{TV}\right)\right) {\rm \,,}
\end{equation}
with $A=A_0/k_{\rm B}$, $B=B_0/k_{\rm B}$ and $V=E\cdot g$, where $k_{\rm B}$ is the Boltzmann constant. 
Equation \ref{G} unlights the dependency of the gas gain versus pressure ($P$), temperature ($T$) and amplification gap thickness ($g$). Those dependencies can be derived from $\Delta G/G = C_P\Delta P + C_T\Delta T + C_g\Delta g$ and expressed as: 
\begin{eqnarray}
\label{Cp}
C_P &=& \frac{1}{G}\frac{\partial G}{\partial P} 
	= \left(\frac{Ag}{T} - \frac{ABPg^2}{T^2V}\right)exp\left(-\frac{BPg}{TV}\right)\\
\label{Ct}
C_T &=& \frac{1}{G}\frac{\partial G}{\partial T} 
	= \left(\frac{ABP^2g^2}{T^3V} - \frac{APg}{T^2}\right)exp\left(-\frac{BPg}{TV}\right)\\
\label{Cg}
C_g &=& \frac{1}{G}\frac{\partial G}{\partial g} 
	= \left(\frac{AP}{T} - \frac{ABP^2g}{T^2V}\right)exp\left(-\frac{BPg}{TV}\right)  {\rm \,.}
\end{eqnarray}
In practice, it is convenient to apply one single correction for pressure and temperature variations using the coefficient:
\begin{eqnarray}
\label{Cx}
C_{P/T} &=& \frac{1}{G}\frac{\partial G}{\partial (P/T)} 
	= \left(Ag - \frac{ABPg^2}{TV}\right)exp\left(-\frac{BPg}{TV}\right)  {\rm \,.}
\end{eqnarray}
 A correction is applied by multiplying the gain by the correction factor $f_x$ given by: 
\begin{eqnarray}
\label{CF}
f_x &=& 1-C_{x} \cdot\Delta \left(x\right)  {\rm\,,}
\end{eqnarray}
where $x$ stands for $g$, $P$, $T$ or $P/T$.

\subsubsection{Application to \ariso and \arco}
The gain dependencies on \emph{P}, \emph{T}, \emph{P/T} and \emph{g} can be predicted from a gain curve by adjusting the constants $A$ and $B$ on the measured trend \emph{via} formula \ref{G}. 
The fits gave $A=(0.14\pm0.01)\degk\mbarinv\muminv$ and $B=(1.8\pm0.1)\degk\volt\mbarinv\muminv$ in \ariso and  $A=(0.10\pm0.01)\degk\mbarinv\muminv$ and $B=(2.1\pm0.2)\degk\volt\mbarinv\muminv$ in \arco.
 The dependencies calculated using the formulae \ref{Cp} -- \ref{Cx} are gathered in table \ref{ArIsoResults}. 
\label{XRayResults}
\begin{table}[h]
	\centering
		\caption{Coefficients predicted from the gain curves in \arisoProp and \arcoProp}
		\begin{tabular}{|c|c|c|c|c|}
				\hline
				Gas			&	$C_P$	(\%\mbarinv)			 &		$C_T$ (\%\degkinv)		&	 	$C_g$	(\%\muminv)			&	$C_{P/T}$ (\%\degk\mbarinv) \\
				\hline
				\hline
				\ariso  &$-0.8 \pm0.08$ & $2.6 \pm0.3$ 	& $-6.6 \pm0.6$	  & $-236 \pm24$\\
				\hline
				\arco		&$-0.5 \pm0.1$ & $1.6 \pm0.4$ & 		$-3.8\pm0.9$ 			& $-145\pm35$\\
				\hline
		\end{tabular}
	\label{ArIsoResults}
\end{table}

\noindent
The predicted values of $C_P$, $C_T$ and $C_{P/T}$ in \arco are compared to direct measurements in the next section.

\subsection{Environmental Study in Ar/CO$_2$ (80/20)}
 \label{envstudy}
\subsubsection{Experimental Conditions}
\label{cond}
During two weeks the amplitude of some 10$^{8}$ pulses from $^{55}$Fe quanta conversions in Ar/CO$_{2}$ 80/20 were recorded, enabling a precise monitoring of the detector gain as a function of time. In parallel, gas pressure and temperature were also recorded. 
The mesh voltage was set to 570$\volt$ at which a gain of about 5$\cdot$10$^{3}$ was measured (see figure \ref{GAINCURVES}). 
The drift field was kept at 100$\voltcm$.
The Ar and CO$_{2}$ gas flows were equal to 0.97 and 0.24$\lh$, yielding a total flow of 1.21$\lh$.  
The mean pressure was 959.7$\mbar$ and the mean temperature was 298.2$\degk$. The temperature was controlled with the help of an air conditioner and the gas pressure fluctuated according to the atmospheric pressure variations.


\subsubsection{Pressure Corrections}
\label{pres}
\begin{figure}[!h]
	\centering
		\includegraphics[width=0.65\textwidth]{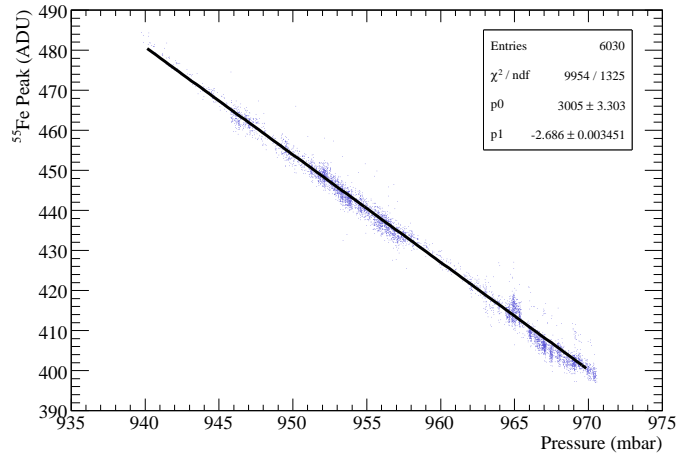}
	\caption{$^{55}$Fe peak value versus atmospheric pressure at constant temperature (T=298\degk).}
	\label{fig:peak_pres_Tcst}
\end{figure}

During part of the run, the gas temperature was maintained around 298$\degk$ to examine the gas gain dependency on pressure only. Figure \ref{fig:peak_pres_Tcst} shows the $^{55}$Fe peak value versus pressure recorded at a temperature $T = (298.0\pm0.5)\degk$. 
A linear behavior is observed and fitted with a slope $\alpha_P$ which relates to the $C_P$ coefficient according to:
\[\alpha_P = \bar{v} \cdot C_P{\rm\,,}\]
where $\bar{v}$ is the average $^{55}$Fe peak value over the fitted range. 
With $\alpha_P = (-2.686\pm0.004)\mbarinv\adu^{-1}$ and $\bar{v}\approx440\adu$, computation gives: 
\[C_P = (-0.61\pm0.01)\%\mbarinv{\rm\ ,}\] 
which is consistent with the predicted value (section \ref{cond}).

\subsubsection{Temperature Corrections}
\label{temp}
\begin{figure}[!h]
	\centering
		\includegraphics[width=0.65\textwidth]{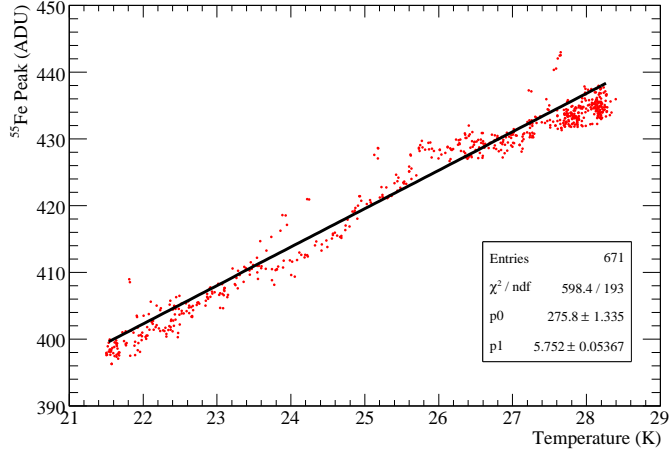}
	\caption{Pressure corrected $^{55}$Fe peak value versus temperature.}
	\label{fig:peak_temp_Pcorr}
\end{figure}

Data recorded during a period with temperature variations of a few kelvins have been corrected for pressure variations using $C_P$ from section \ref{pres} and formula \ref{CF}. The corrected $^{55}$Fe peak value, $v_{{\rm corr}_P}$, is given by:
\begin{eqnarray}
v_{{\rm corr}_P} = v \cdot \left(1 - C_P \cdot \Delta P\right) {\rm \,,}
\end{eqnarray}
where $v$ is the raw peak value, and is plotted in figure \ref{fig:peak_temp_Pcorr}. 
 A linear fit was performed and its slope $\alpha_T$ gave the $C_T$ coefficient through 
$\alpha_T = \bar{v} \cdot C_T$. With $\alpha_T = 5.75 \degkinv\adu^{-1}$ and $\bar{v}\approx420$, computation leads to: 
\[C_T = (-1.37\pm0.01)\%\degkinv{\rm\ ,}\] 
which is consistent with the value predicted in section \ref{cond}. 

\subsubsection{Corrections using the Ratio of Pressure over Temperature}
\label{PoverT}
\begin{figure}[!h]
	\centering
		\includegraphics[width=0.65\textwidth]{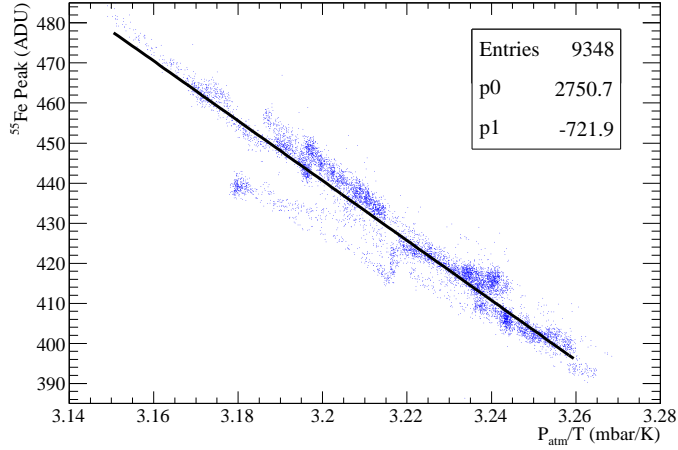}
	\caption{$^{55}$Fe peak value versus the ratio of pressure over temperature, using all data without any correction.}
	\label{fig:peak_Pres-temp}
\end{figure}

The evolution of the $^{55}$Fe peak value along the whole data set versus the ratio of pressure over temperature is plotted in figure \ref{fig:peak_Pres-temp}. 
A straight line was adjusted on the points and its slope $\alpha_{P/T}$ gave the $C_{P/T}$ coefficient through 
$\alpha_{P/T} = \bar{v} \cdot C_{P/T}$. With $\alpha_{P/T} = -722 \degkinv\adu^{-1}$ and $\bar{v}\approx440$, its value is:
 \[C_{P/T} = (-164\pm1)\%\degk\mbarinv{\rm\ ,}\]
 which is within the error range of the value predicted in section \ref{cond}.

\subsubsection{Conclusion of the Study}
\label{XRAYconclusion}
Direct measurement of the coefficients $C_P$, $C_T$ and $C_{P/T}$ showed good agreement with the gas gain model prediction (table \ref{ArCO2CompPredMeas}). 
Using those coefficients, the $^{55}$Fe peak value has been corrected for pressure and temperature variations according to: 
\begin{eqnarray}
v_{{\rm corr}_P} &=& v \cdot \left(1 - C_P \cdot \Delta P\right)\\
v_{{\rm corr}_T} &=& v \cdot \left(1 - C_P \cdot \Delta P\right)\cdot \left(1 - C_T \cdot \Delta T\right)\\
v_{{\rm corr}_{P/T}} &=& v \cdot \left(1 - C_{P/T} \cdot \Delta \left(P/T\right)\right) {\rm \,.}
\end{eqnarray}
\begin{table}[h]
	\centering
		\caption{Summary of predicted and measured values for environmental coefficients in Ar/CO$_2$(80/20)}
		\begin{tabular}{|c|c|c|c|}
				\hline
				Coefficient &$C_P	$										 &			$C_T	$							&	$C_{P/T}$ \\
				\hline
				Predicted Value &$(-0.5 \pm0.1)\%\mbarinv$ & $(1.6 \pm0.4)\%\degkinv$ & $(-145\pm35)\%\degk\mbarinv$\\
				\hline
				Measured Value  &$(-0.61\pm0.01)\%\mbarinv$ & $(1.37\pm0.01)\%\degkinv$ & $(-164\pm1)\%\degk\mbarinv$\\
				\hline
		\end{tabular}
	\label{ArCO2CompPredMeas}
\end{table}

\begin{figure}[!h]
	\centering
		\includegraphics[width=0.80\textwidth]{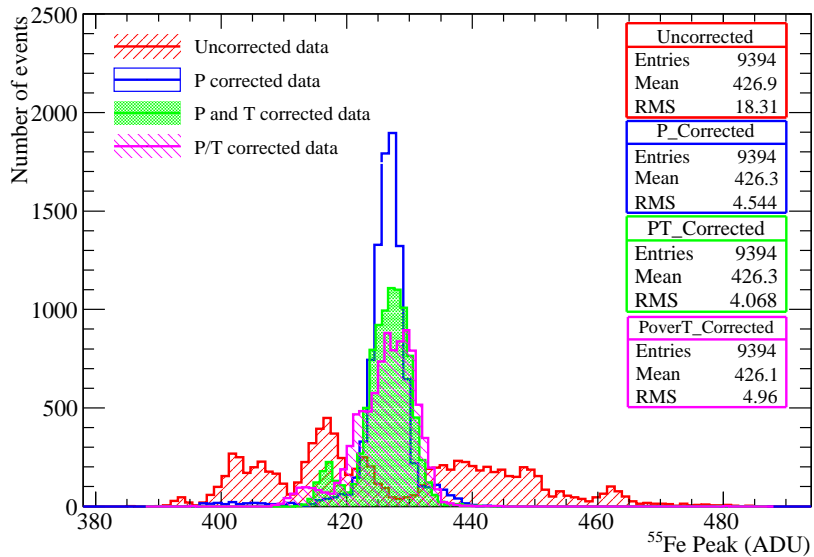}
	\caption{Summary of the corrections applied to the data.}
	\label{fig:peakScat_allCase-2}
\end{figure}

The result of those corrections are gathered in figure \ref{fig:peakScat_allCase-2}. The $^{55}$Fe peak value is very scattered before applying any correction. The successive corrections for pressure and temperature leads to a major improvement of the $^{55}$Fe peak value regularity. The direct correction using the ratio of pressure over temperature is also valuable. 
The correction yielding the strongest reduction of the distribution r.m.s. is the one based on C$_P$ because the temperature was controlled during the data acquisition and showed limited variations.

Those results validate the gas gain model and the method for environmental corrections of the data.


\subsection{X-Ray Test Conclusion}
\label{xrayconc}
The mesh collection efficiency should be maximum for a field ratio around 200 in \ariso. 
Hence for later measurements, as the mesh voltage will be around -420\volt, the cathode voltage will be kept 50\volt below. 

In a given gas mixture, the gas gain depends primarily on the gas number density and the amplification gap thickness.
The density relates to the ratio P/T which, in Ar/C$_4$H$_{10}$ (95/5), impacts on the gain according to -236\%\degk\mbarinv.  
Typical values of this ratio are around $3\cdot10^{-2}\mbar\degkinv$ with variations of the order of 10$^{-2}$\mbar\degkinv leading to some 2 -- 3\% gas gain fluctuation. 
If the correction factor is mainly below 10\%, the error margins make the corrections uncertain so they shall be applied only if a significant amount of data need a correction factor above 10\%. 

The amplification gap size determines the distance over which an electron avalanche develops and the amplification field for a given mesh voltage. 
It strongly impacts on the gain.  
 The bulk planarity is better than 5\mum, but a gap variation of 1\mum should result in a change of the gas gain of 6.6\% in Ar/C$_4$H$_{10}$ (95/5). Therefore the mesh irregularities are expected to play a major role in the detector's gain disparity. 
Smaller variations are predicted in Ar/CO$_2$ (80/20) due to a milder dependence of the first Townsend coefficient on the electric field.

\section{Measurements with Particle Beams}

\subsection{Experimental Layout}
\begin{figure}[!h]
	\centering
		\includegraphics[width=1.00\textwidth]{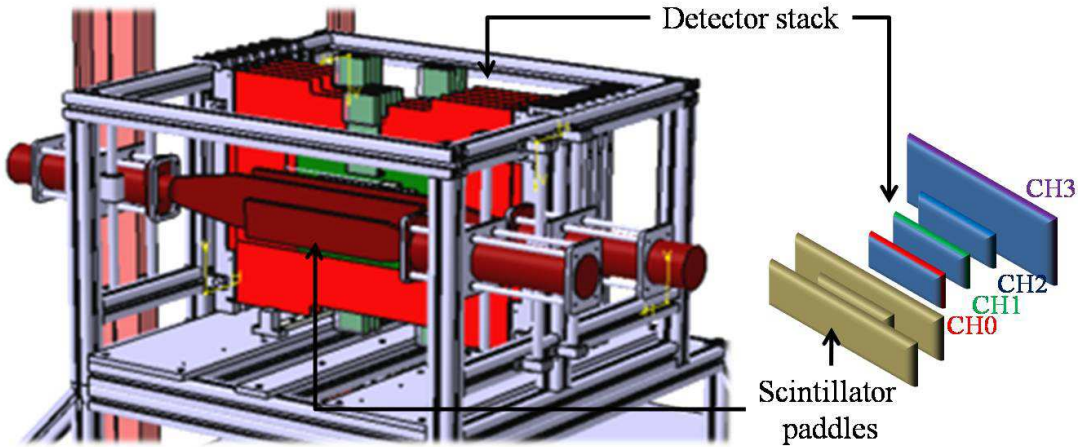}
	\caption{The test beam layout, including detectors, scintillator paddles and mechanical structure (left). A corresponding schematic view (right).}
	\label{TBStruct}
\end{figure}

The detector stack was set up placing the small chambers in the front followed by the large one at the rear (figure \ref{TBStruct}).  
The distance between each chamber was 10$\cm$. 
Three scintillator paddles were placed in front of the stack, the trigger signal was provided by the triple time coincidence of their output. Two of them were 8$\times$32$\cma$ and the last one had the exact dimensions of the small chambers (6$\times$16$\cma$). 

A common pre-mixed \arisoProp gas was used and the voltage applied on the prototypes' meshes were  -420$\volt$, -420$\volt$, -430$\volt$ and -410$\volt$, for CH0 to CH3, respectively. 
These voltage values where set as a trade off between a high gain and a spark rate below about one per hour. 
The drift voltages were set 50$\volt$ below the meshes' ones so that the fields ratios were always corresponding to the maximum collection efficiency (see section \ref{EFFCOLL}).

\subsection{Particle Sources}

\subsubsection{CERN/SPS, H2 beam line}
The Super Proton Synchrotron (SPS) delivers a 400$\gevc$ proton beam for fixed target experiments. 
Data were taken with a secondary 200$\gevc$ negative muon beam. The beam was available during a 9$\sec$ spill period every machine cycle. Each machine cycle lasted for 33$\sec$ during night time and 48$\sec$ during day time. The beam was intense enough to saturate the acquisition rate at about 130 events$/\sec$. Data were recorded during August 2008 beam test session. 

\subsubsection{CERN/PS, T9 beam line}
The Proton Synchrotron (PS) delivers a 28$\gevc$ proton beam for injection in SPS and CERN's East Area's Fixed target experiments. Data were recorded with a secondary 7$\gevc$ positive pion beam. 
The beam was available during one to three 0.4$\sec$ spill periods every machine cycle. A machine cycle lasted for a variable time around 40$\sec$. The beam was also intense enough to saturate the acquisition rate. Data were recorded during November 2008 beam test session.

\subsection{Data}
\subsubsection{Environmental and Noise Conditions}
\begin{figure}[h]
	\centering
		\includegraphics[width=0.495\textwidth]{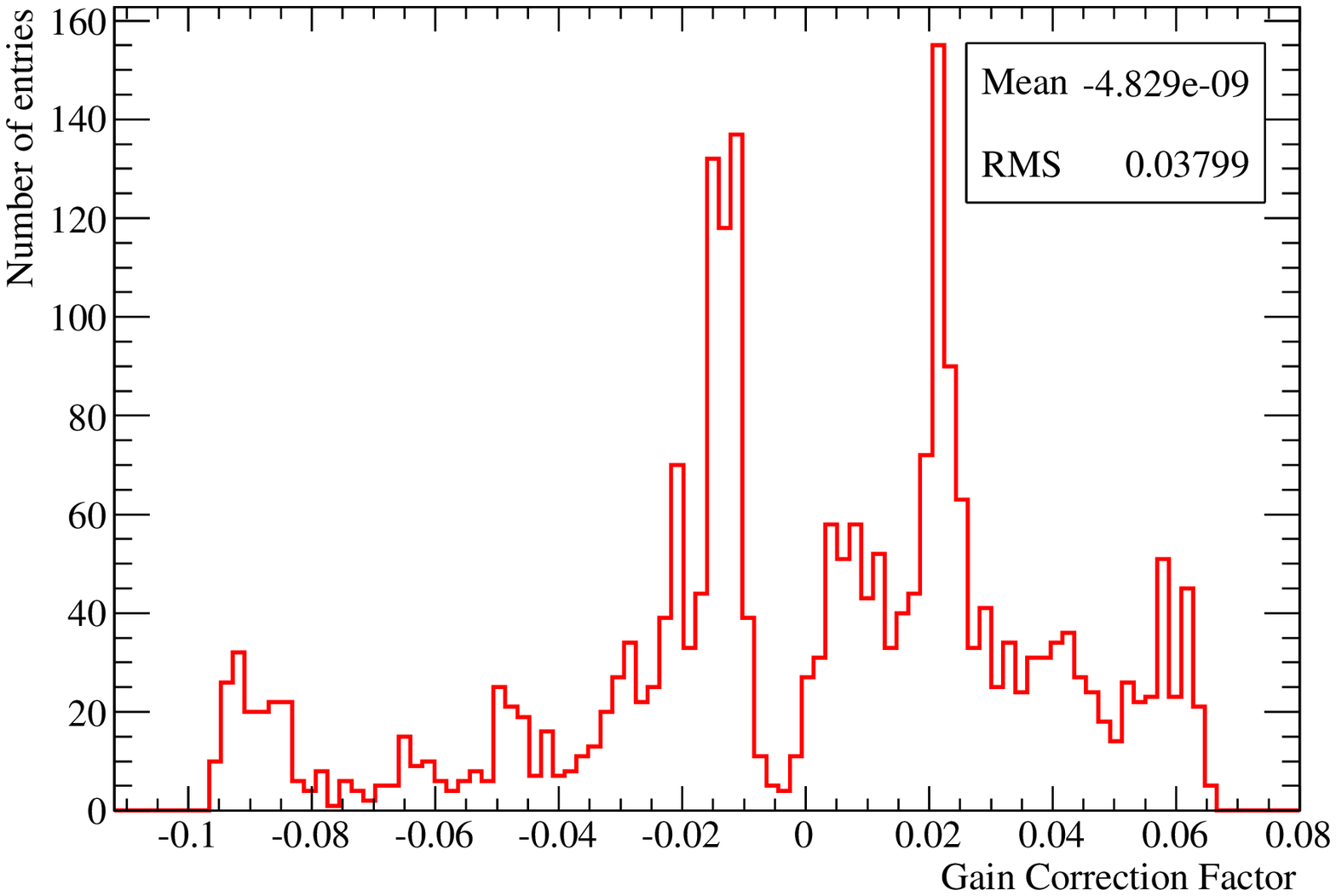}
		\includegraphics[width=0.495\textwidth]{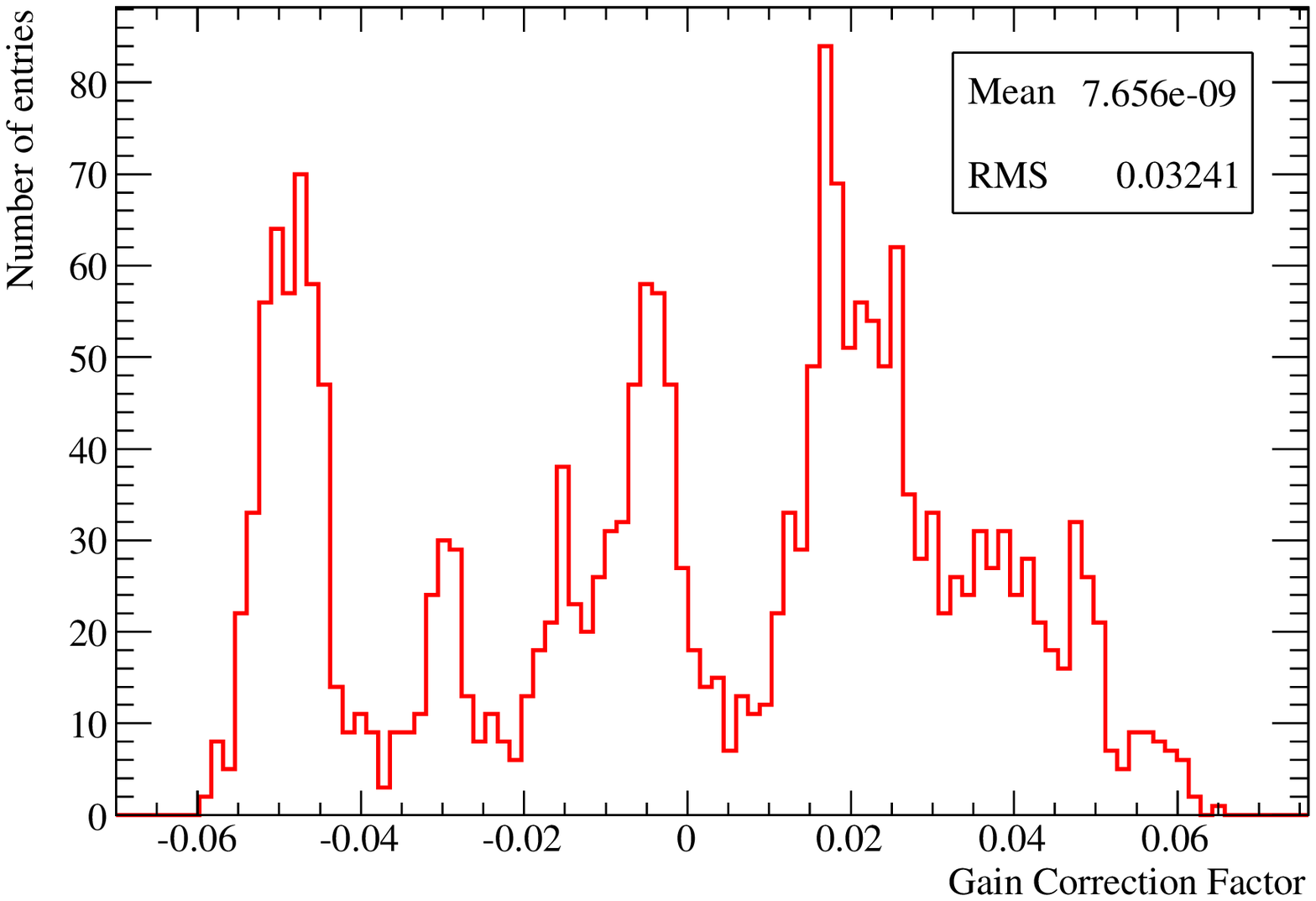}
	\caption{Histograms of the correction factors for data from August (left) and November (right).}
	\label{fig:CorrFactor}
\end{figure}

During the data acquisition, the atmospheric pressure and the gas temperature were monitored. Using $C_{P/T} =-2.36\degk\mbarinv$, a gain correction factor was computed using formula \ref{CF} and found always below 10\% with an r.m.s. below 4\% (see figure \ref{fig:CorrFactor}). In accordance with section \ref{xrayconc} environmental corrections are sufficiently small and considered to be negligible. 
Moreover, in a digital detector those corrections could not be applied at all. Therefore, in the aim of a DHCAL the results given here will remain uncorrected.

\begin{figure}[!h]
	\centering
		\includegraphics[width=\textwidth]{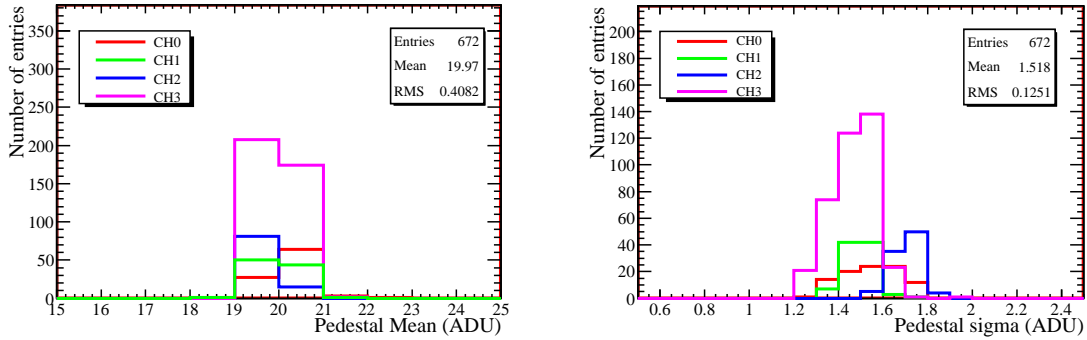}
	\caption{Pedestal Mean (left) and sigma (right) distributions.}
	\label{fig:PedMeanSig}
\end{figure}

The GASSIPLEX pedestals were periodically aligned at 20$\adu$ on the V550 ADC modules. 
They were measured to be at this value with 2\% r.m.s. variations over all channels through the whole data set (see figure \ref{fig:PedMeanSig}, left). The pedestal sigmas were obtained from a gaussian fit and showed an average value of 1.5$\adu$ corresponding to 0.3$\fc$ or 2000 $\en$ (figure \ref{fig:PedMeanSig}, right). These figures demonstrate very good noise conditions.

\subsubsection{Event tags}
All channels were recorded without threshold.
A hit is defined by applying an off-line threshold equal to 1.5$\fc$ (7$\adu$ above pedestal).
Two types of events were selected for the analysis: the Platinum and the Golden events.
\begin{itemize}
	\item \emph{Platinum events}\label{Pt}: an event is tagged as platinum by requiring one single hit in each of the four chambers.	Those events are used for gain and pedestal studies since they ensure a very low noise hit contamination.
	\item \emph{Golden events}\label{Au}: a golden event is selected by requiring one single hit in three out of the four chambers.
		Those events are used for efficiency and multiplicity studies.
\end{itemize}

\subsection{Gain Distribution Measurement}
\label{GainDistrib}
\begin{figure}[!h]
	\centering
		\includegraphics[width=0.49\textwidth,trim=0 0 45 35,clip]{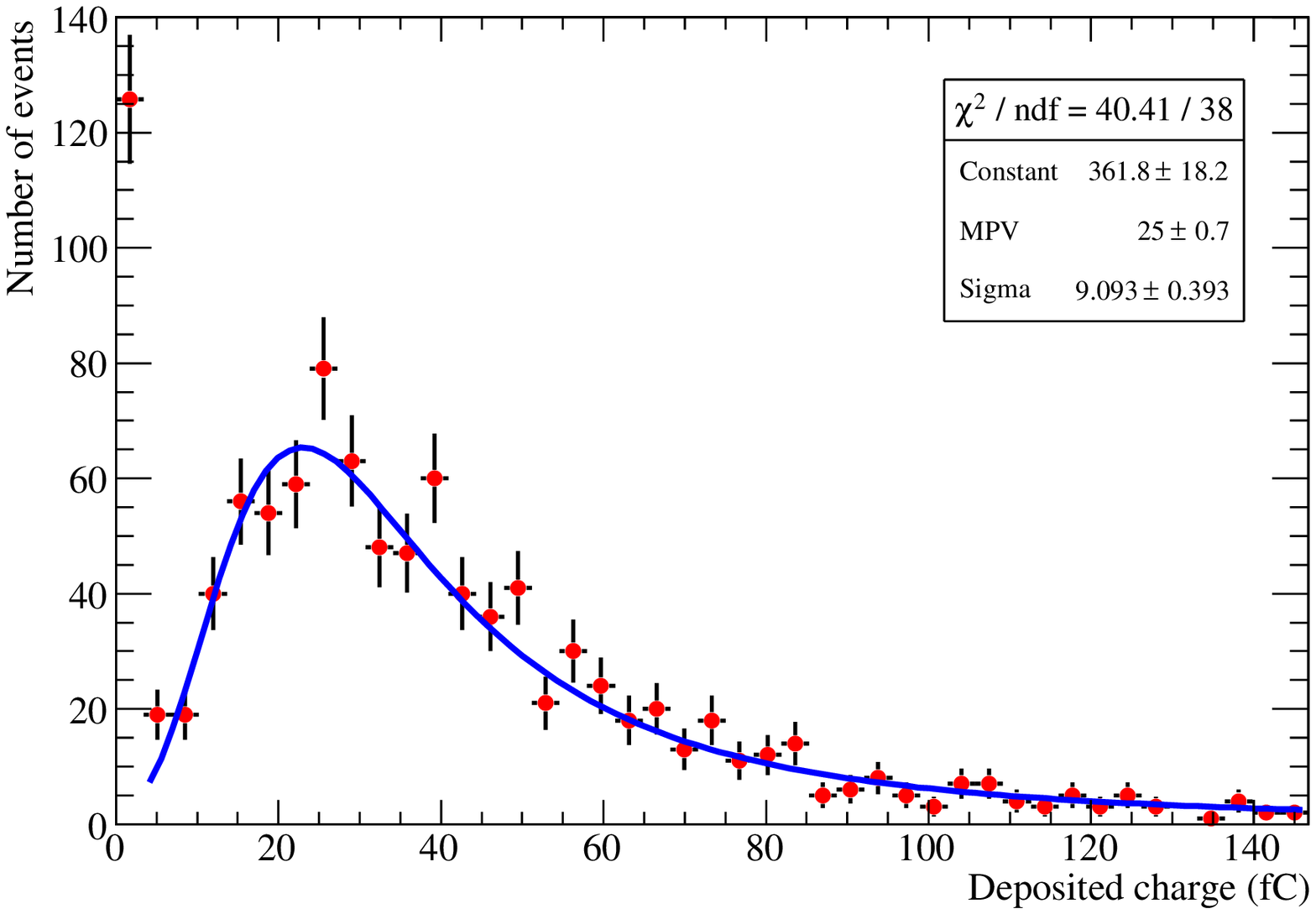}
		\includegraphics[width=0.49\textwidth,trim=0 0 45 35,clip]{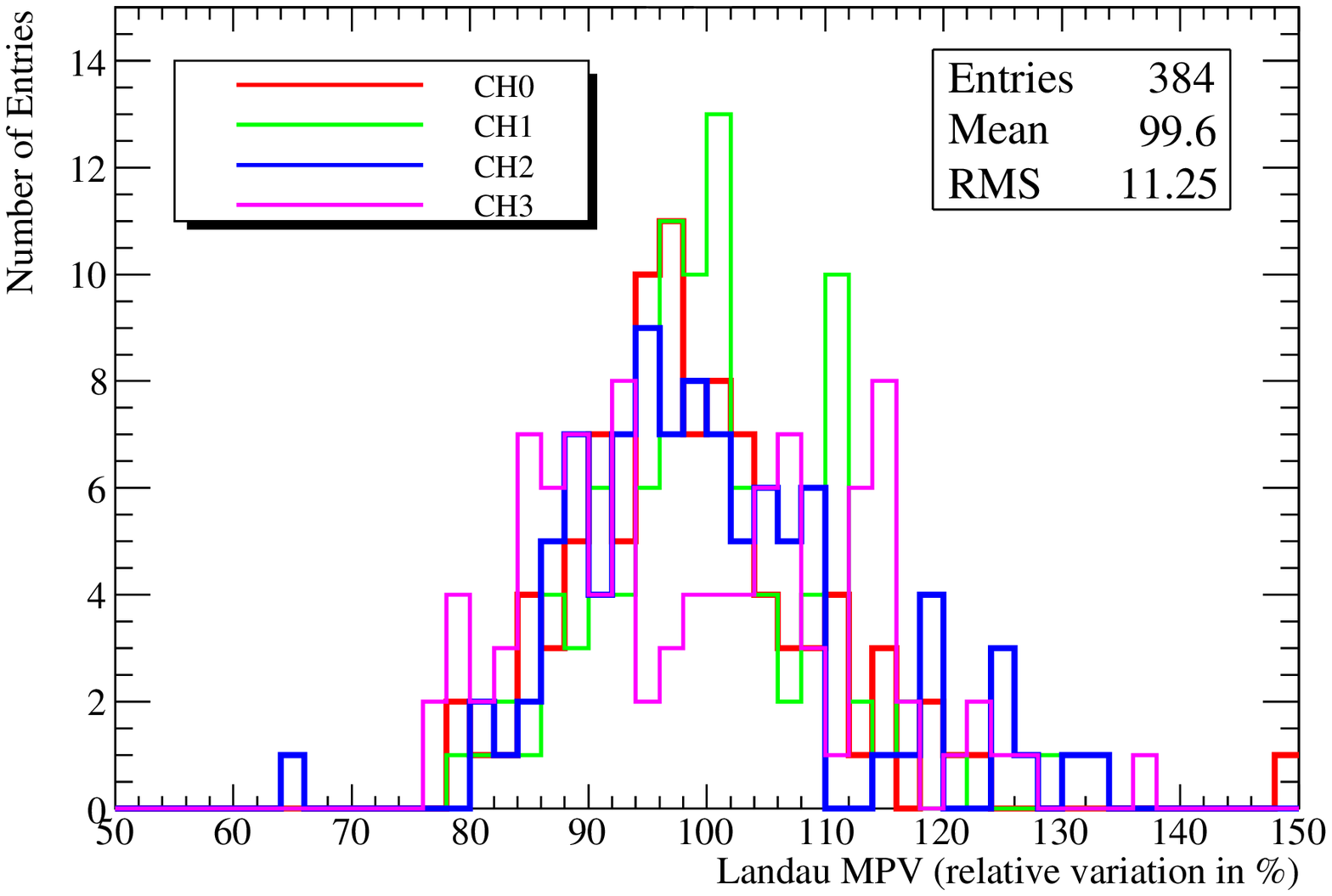}
	\caption{Fit of a Landau function to the data for a single channel, the pedestal has been scaled to fit in the vertical range. The spectrum was built using Platinum events (left). Landau MPV distribution normalized to 100, for all chambers (right).}
	\label{landau}
\end{figure}

\begin{figure}[!h]
	\centering
		\includegraphics[width=1.0\textwidth,clip,trim= 15 5 10 10]{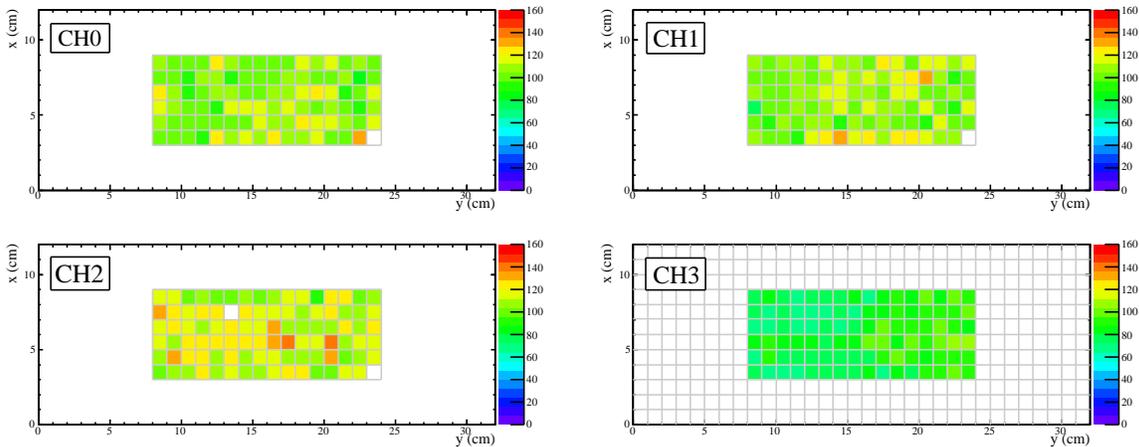}
		\caption{Landau MPV maps of all prototypes (color axis in ADU).}
	\label{fig:MPVmapsMIX}
\end{figure}

For every channel, a Landau function was fitted on the data from platinum events (see figure \ref{landau} (left)) and its Most Probable Value (MPV) was defined as the detector's global gain for charged particles (conversion, mesh amplification and electronics amplification). The resulting values are mapped in figure \ref{fig:MPVmapsMIX}. 
 
The most probable deposited charge, averaged over all channels, is 22.6, 22.9, 24.5 and 17.5\fc for CH0 to CH3 respectively.
The relative gain distribution of all the channels is shown in figure \ref{landau} (right) having an r.m.s. of 11.25\%. 
Since the electronics gain distribution has a very low r.m.s., this value is expected to be mainly due to drift and/or amplification gaps non-uniformity.

\subsection{Efficiency Measurement}
\label{eff}
\begin{figure}[!h]
	\centering
		\includegraphics[width=1.00\textwidth,clip,trim= 15 5 10 10]{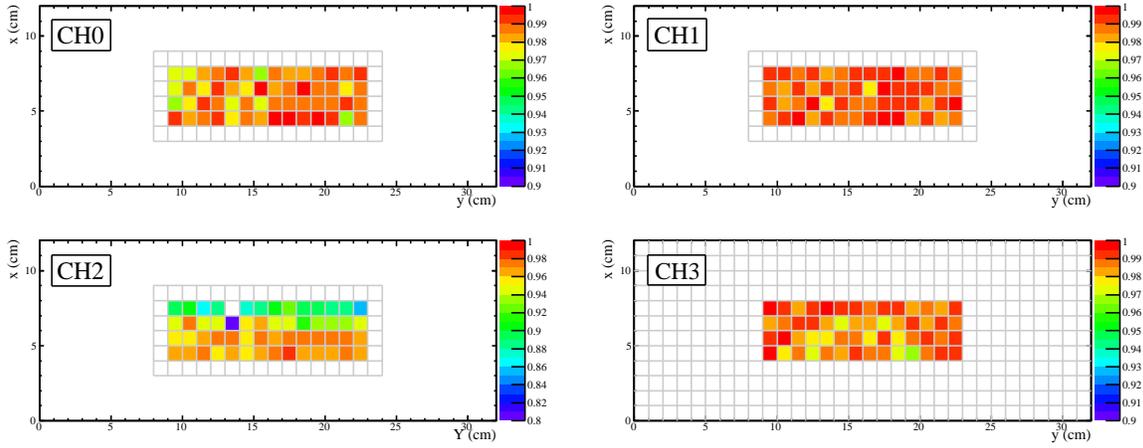}
		\caption{Maps of each prototypes efficiency. The measurement was performed for a fiducial area omitting all border pads and using a 3$\times$3$\pad$ area around the expected hit to avoid misalignment issues.}
	\label{fig:effmaps}
\end{figure}

\begin{figure}[!h]
	\centering
		\includegraphics[width=0.65\textwidth]{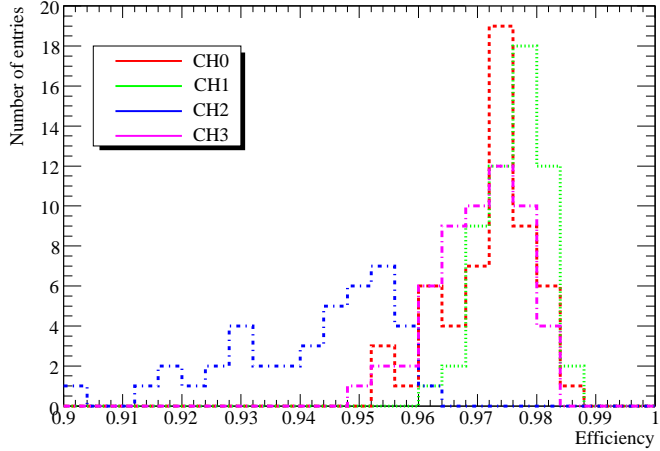}
	\caption{Pad efficiency distribution for each chamber.}
	\label{fig:effhistos}
\end{figure}

In order to measure the efficiency of a given chamber, a sub range of golden events 
was selected by requesting three aligned hits in the three other chambers to define a particle track.  A safety threshold of 12.5$\fc$ was applied for the three reference chambers' hits to completely avoid taking noise hits into account. 
In each processed event, a hit was searched in a 3$\times$3$\pad$ area centered at the intersection between the extrapolated reconstructed particle track and the chamber plane.
The resulting efficiencies are mapped in figure \ref{fig:effmaps} and their distribution is shown in figure \ref{fig:effhistos}.

The hit background was estimated  by counting the number of hits outside the 3$\times$3$\pad$ area. This number, normalized to the 3$\times$3$\pad$ area,  was subtracted to measure the final efficiency (Table \ref{tab:EfficiencyMeasurements}). 
Thanks to the very low threshold, three chambers show an excellent efficiency, larger than 97\%.
CH2 shows a lower efficiency (91\%). It might be due to the lower tension of its mesh or to the broader pedestals of its electronics.

\begin{figure}[!h]
	\centering
		\includegraphics[width=0.65\textwidth]{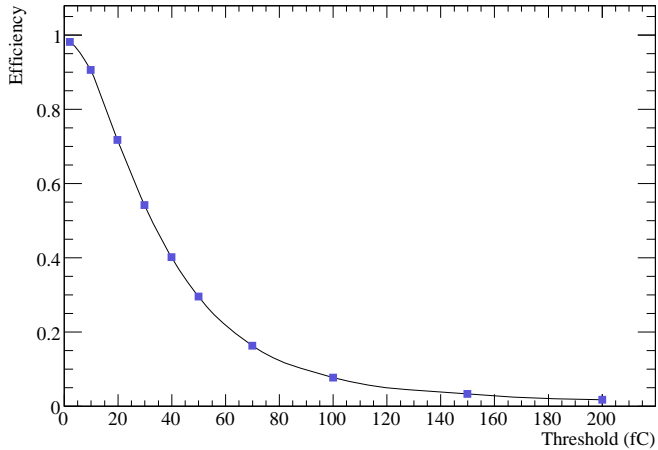}
	\caption{Efficiency versus threshold (CH1).}
	\label{fig:EffvsThreshold}
\end{figure}

\begin{table}[!h]
	\centering
			\caption{Efficiency measurements for a 1.5$\fc$ threshold.}
			\begin{tabular}{|c|c|c|c|}
		\hline
			Chamber & Raw efficiency & Noise hit fraction 		& Noise corrected efficiency \\\hline\hline
			CH0			& (99.0$\pm$0.1)\%	 & (1.3$\pm$0.1)\%	& (97.7$\pm$0.1)\% 	\\\hline
			CH1			& (99.0$\pm$0.1)\% 	 & (1.3$\pm$0.1)\%	& (97.7$\pm$0.1)\%	\\\hline
			CH2			& (93.0$\pm$0.1)\% 	 & (2.0$\pm$0.1)\%	& (91.2$\pm$0.1)\%	\\\hline
			CH3			& (98.8$\pm$0.1)\% 	 & (1.6$\pm$0.1)\%	& (97.2$\pm$0.1)\%	\\\hline
		\end{tabular}
	\label{tab:EfficiencyMeasurements}
\end{table}
In a DHCAL, such a low threshold (1.5\fc) may not be achievable. 
Thus the same study was carried out for each chamber varying the threshold from 1.5$\fc$ to 200$\fc$. The dependency between efficiency and threshold is unlighted in figure \ref{fig:EffvsThreshold}.  
A steep drop of efficiency with threshold is observed. The efficiency is about 70\% at a threshold of 20\fc and drops below 10\% for thresholds higher than 100\fc. This has strong concequences on the electronics noise requirements.

\subsection{Multiplicity Measurement}
Still using golden events, the number of hits in a 3$\times$3$\pad$ area around the pad expected to be hit was counted.
The multiplicity is computed as the mean of this number over all the processed events (Table \ref{tab:mult}) is corrected with the same method as for the efficiency (section \ref{eff}). 
\begin{table}[!h]
	\centering
			\caption{Multiplicity measurements for a 1.5$\fc$ threshold.}
		\begin{tabular}{|c|c|c|}
		\hline
 Chamber	& Raw multiplicity & Noise corrected multiplicity\\	\hline\hline
	CH0				&			1.070$\pm$0.008			&			1.057$\pm$0.008	\\
	CH1				&			1.080$\pm$0.008			&			1.065$\pm$0.008	\\
	CH2				&			1.090$\pm$0.008			&			1.070$\pm$0.008	\\
	CH3				&			1.114$\pm$0.008			&			1.096$\pm$0.008	\\	\hline
		\end{tabular}
	\label{tab:mult}
\end{table}

\begin{figure}[!h]
	\centering
		\includegraphics[width=0.65\textwidth]{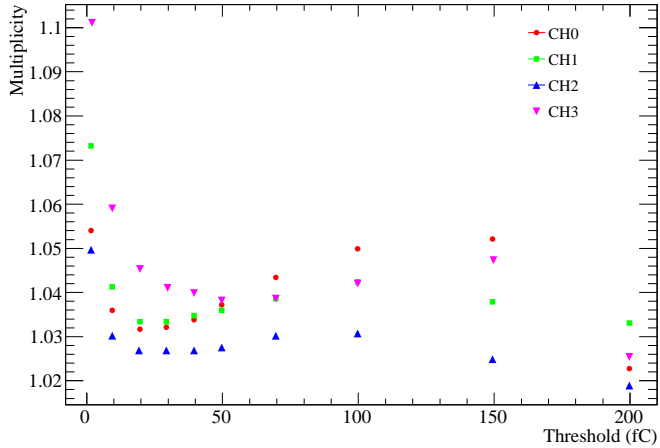}
	\caption{Multiplicity versus threshold for the four chambers.}
	\label{fig:MVSTallCH-aug}
\end{figure}

The four chambers showed a noise corrected multiplicity between 1.06 and 1.10, which is a benefit for a PFA algorithm. 
The behavior of multiplicity versus threshold was also studied and is illustrated in figure \ref{fig:MVSTallCH-aug}. 

After a quick fall, the multiplicity rises slowly and then decreases at high threshold. At very low threshold, almost all pads receiving charge are seen and the maximum multiplicity is measured. With increasing  threshold the multiplicity decreases dramatically down to 1.03 -- 1.04 at 30 -- 40\fc. Above this value, mainly events of particles with large energy deposit are considered. These particles likely produce $\delta$-rays leading to some ionization far from the track and hence to a higher multiplicity. Above 150\fc, as a consequence of the decreasing detection efficiency with threshold, the multiplicity declines again as expected. 


As a conclusion, the efficiency and hit multiplicity of the MICROMEGAS prototypes are potentially excellent (97\% efficiency and 1.06 multiplicity at a 1.5\fc threshold) according to the requirements of a DHACL active layer. More technical details about the beam test analysis are available in \cite{AE}. 
%

\section{Embedded Digital Readout Prototype}
A compact detector is compulsory for the DHCAL active layer.
Therefore, a process has been developed to reach a bulk MICROMEGAS with embedded readout chip. The DIRAC chip has been chosen for the first prototype. This was also an opportunity to test this R\&D chip in real conditions.

\subsection{DIRAC ASIC}
DIRAC \cite{DIRAC} is a 64-channel digital ASIC intended for the readout of gaseous detectors like MICROMEGAS, GEM, GRPC. Its design is highly specific to ILC DHCAL requirements. It is based on low cost technologies, offers a low power consumption thanks to power pulsing synchronized to the ILC clock. It provides two operative modes (high gain in MICROMEGAS/GEM mode and low gain in RPC mode), a scale of three thresholds with an 8-bit precision and a fully digital serial interface.

\subsection{Prototype Layout}
The prototype consists of an 8$\times$8$\pad$, 6 layers, class 6 PCB, equipped with a single DIRAC ASIC. A mask was fixed on the PCB's ASIC side in order to avoid the embedded electronics from being squashed during the lamination of the mesh.
The same bulk layout as the one described in section \ref{proto} was used. The anode segmentation was made of 1$\times$1\cma pads spaced every 500\mum.

\subsection{Beam Test Result}
\begin{figure}[!h]
	\centering
		\includegraphics[width=0.40\textwidth,bb=0 0 550 710,angle=270,trim= 120 0 0 0,clip]{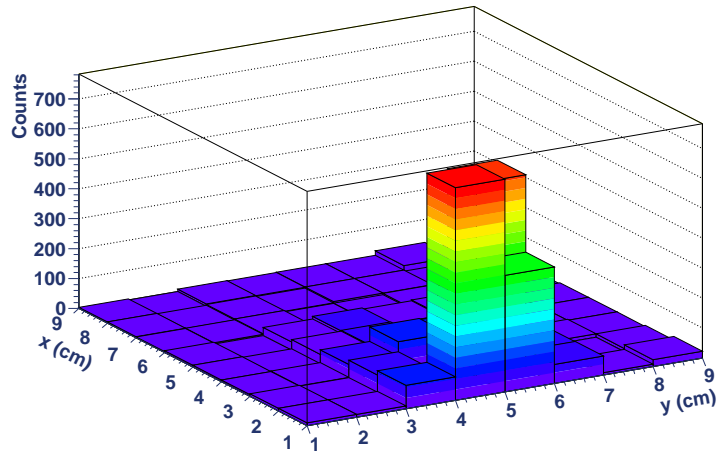}
	\caption{Beam profile obtained with digital readout using the DIRAC ASIC.}
	\label{DIRACbp}
\end{figure}

The first operative test with bulk MICROMEGAS with embedded electronics has been carried out and the electronics proved to have survived the lamination process by showing the beam profile displayed in figure \ref{DIRACbp}. 
This profile was obtained in a 200\gevc muon beam, with a  19\fc threshold, in \arisoProp, at a mesh voltage of 410\volt and a drift voltage of 460\volt. The data acquisition was auto-triggered by the ASIC.
A raw hit multiplicity of 1.1 has been determined. Further measurements will be performed with a stack of several embedded DIRAC MICROMEGAS detectors.
This test is a proof of principle for the bulk MICROMEGAS with embedded electronics and for the DIRAC ASIC functionality.

\section{Conclusion}
\begin{sloppypar}
Bulk MICROMEGAS prototypes with analog readout have been tested with X-rays and CERN's particle beams.

The gas gain dependency on pressure, temperature and amplification gap thickness variations has been calculated from gain curves in \arisoProp and \arcoProp. As expected from the steeper gain curve, the values obtained in the Ar/iC$_4$H$_{10}$ mixture are significantly higher. This difference makes Ar/CO$_2$ more stable against pressure and temperature variations than Ar/iC$_4$H$_{10}$ making this gas mixture interesting despite the lower gain it provides. In \arcoProp, these calculated values were confronted with measurements showing good agreement. A method for gain correction based on those dependencies has been presented.

Four chamber were placed in 200\gevc muon and 7\gevc pion beams. The gain distribution of 384 channels (a 384\cma area) showed an 11\% r.m.s. variation.
The efficiency and the hit multiplicity were measured and their dependency versus threshold was studied.
An efficiency near 97\% at a 1.5$\fc$ threshold was obtained and a hit multiplicity as low as 1.03 has been found at 20\fc.

The first bulk MICROMEGAS with embedded readout electronics have been built, tested and proved to be functional, which  validates the fabrication process of a compact MICROMEGAS and also the DIRAC ASIC performances.

\end{sloppypar}

\acknowledgments
We would like to thank Bruno Chauchaix, Lau Gatignon and Dragoslav-Laza Lazic for their very precious help while on the test areas. We are also grateful to Didier Roy for his hard work on the CENTAURE software for data acquisition and online monitoring. We wish to thank Rui de Olivera and the CERN TS-DEM group for the bulk production and their collaboration in the development of the prototype with embedded electronics. We also thank Philippe Abbon for the production of the 613V GASSIPLEX boards. A special thank also for Kostas Karakostas, Glenn Cougoulat, Richard Hermel
and
Jean Tassan
 for their contribution at the begining of the project and  Denis Fougeron for designing the analog prototype PCBs.

\end{document}